\documentclass[11pt,a4paper]{article}
\usepackage{graphicx}
\newcommand {\be} {\begin{eqnarray*}}
\newcommand {\ee} {\end{eqnarray*}}
\newcommand {\bea} {\begin{eqnarray}}
\newcommand {\eea} {\end{eqnarray}}

\begin{document}

\begin{center}
{\Large{\textbf{Astrophysical Implications of Higher-Dimensional Gravity}}}
\end{center}    

\begin{center}
\textbf{Tom$\acute{\mbox{a}}\check{\mbox{s}}$ Liko$^{\dagger}$, James M. Overduin$^{\ddagger}$, and Paul S. Wesson$^{\dagger}$} \\
{\small $^{\dagger}$Department of Physics, University of Waterloo} \\
{\small Waterloo, Ontario, Canada, N2L 3G1} \\
{\small $^{\ddagger}$Astrophysics and Cosmology Group, Department of Physics,
Waseda University} \\
{\small Okubo 3-4-1, Shinjuku-ku, Tokyo 169-8555, Japan}
\end{center}

\begin{center}
Key words:\\
Dark matter, Cosmological constant, Microwave background,
Tests, Big Bang, Extra dimensions of spacetime
\end{center}

E-mail addresses: 1. tliko@astro.uwaterloo.ca\\
\phantom{aaaaaaaaaaaaaaaaaaa}2. overduin@gravity.phys.waseda.ac.jp\\
\phantom{aaaaaaaaaaaaaaaaaaa}3. wesson@astro.uwaterloo.ca

\begin{abstract}
We review the implications of modern higher-dimensional theories of gravity for
astrophysics and cosmology.  In particular, we discuss the latest developments of
space-time-matter
theory in connection with dark matter, particle dynamics and the cosmological constant,
as well as related aspects of quantum theory.  There are also more immediate tests of
extra dimensions, notably involving perturbations of the cosmic 3K microwave background
and the precession of a supercooled gyroscope in Earth orbit.  We also outline some
general features of embeddings, and include pictures of the big bang as viewed from a
higher dimension.
\end{abstract}
\pagebreak
\tableofcontents
\section{Introduction}
The gravitational theories that we present here are built on the bedrock of
general relativity.  They are minimal extensions of Einstein's theory,
in the sense that they are based on the same principle (general covariance),
derived from the same action, characterized by the same field equations, and expressed
in the same mathematical language (the language of tensors).  Only the vocabulary
of this language is extended: tensors are allowed to depend on more than four
coordinates.  While this step is simple conceptually, its implications for fields
such as astrophysics and cosmology are far-reaching and diverse.  The implications
may prove to be comparable to the changes that followed the replacement of
Newton's absolute time and three-dimensional space with the four-dimensional (4D)
spacetime of Minkowski.

We begin by laying out the main equations of general relativity (Weinberg 1972),
which will be needed below.  Prime among these are Einstein's field equations
\bea
R_{\alpha\beta} - \frac{1}{2}Rg_{\alpha\beta} + \Lambda g_{\alpha\beta} 
= \frac{8\pi G}{c^4}T_{\alpha\beta}.
\label{geinstein}
\eea
Here the metric tensor $g_{\alpha\beta}$ relates distances in space and time via
$ds^2=g_{\alpha\beta}dx^{\alpha}dx^{\beta}$ and may be seen as a generalization
of the Pythagorean theorem to curved 4D spacetime.  The Ricci tensor $R_{\alpha\beta}$
encodes information about the curvature, and its contraction
$R=g^{\alpha\beta}R_{\alpha\beta}$ is the Ricci scalar.
The material contents of the system, whatever it might be, are described by the
stress-energy tensor $T_{\alpha\beta}$ on the right-hand side of the equations.
The quantity $\Lambda$ is the cosmological constant, which manifests itself as an
acceleration whose magnitude increases with distance $r$, i.e. $|\Lambda|c^{2}r/3$
in the weak-field limit.  The dimensions carried by $\Lambda$ are $L^{-2}$, which
implies that this term must be significant globally.

Taking the trace of equation~(\ref{geinstein}) gives $R=4\Lambda-(8\pi G/c^4)T$ where
$T=g^{\alpha\beta}T_{\alpha\beta}$, so that these equations may be rewritten as
\bea
R_{\alpha\beta} = \frac{8\pi G}{c^4}\left( T_{\alpha\beta} - \frac{1}{2} T g_{\alpha\beta}
   \right) + \Lambda g_{\alpha\beta},
\label{geinstein2}
\eea
where the cosmological constant is now treated as a source-term along with the
stress-energy tensor of matter.  In vacuum, where there is no matter ($T_{\alpha\beta}=0$),
these equations reduce to
\bea
R_{\alpha\beta} = \Lambda g_{\alpha\beta}.
\label{vacuumspacetime}
\eea
The cosmological constant $\Lambda$ thus specifies the curvature associated with empty
spacetime.  The ten equations contained in (\ref{vacuumspacetime}) serve in principle to
determine the ten potentials contained in $g_{\alpha\beta}$.  Well known solutions of
these equations include those of Schwarzschild, de~Sitter, Milne and Kerr.

In higher-dimensional (or Kaluza-Klein) gravity, one retains the formalism above but allows
the tensor indices to range over $0,\ldots,N-1$ rather than $0,\ldots,3$.  We will
concentrate here on the case $N=5$ (Overduin and Wesson 1997; Wesson 1999).  A question
that arises immediately is why the fifth dimension does not make its presence obvious at
low energies.  Kaluza's original approach to this problem was to set all derivatives with
respect to $\ell$ to zero (the cylinder condition), and this was supported shortly afterward
by Klein who showed that it would follow if the topology of the extra coordinate were 
closed with a \emph{tiny} radius of curvature (on the order of the Planck length).  The
momenta of all particle states in the direction associated with this coordinate are then
pushed to Planck energies, far beyond the reach of experiment.  This approach has
subsequently been extended to include the weak and strong interactions as well.  In this
way, Kaluza and Klein laid the foundations for modern unified-field theories such as 10D
superstrings (Green et al. 1987) and 11D supergravity (Buchbinder and Kuzenko 1998).  The
value of $N(\geq4)$ is, however, unknown.

Following the work of Einstein in 4D field theory, if the form of the 5D line element
$dS^{2}=g_{AB}dx^{A}dx^{B}$ is known, then the 5D Ricci tensor $R_{AB}$ and Ricci scalar
$R=g^{AB}R_{AB}$ can be constructed in the same way as their 4D counterparts, and one
obtains field equations exactly the same as those in (\ref{geinstein2}) and 
(\ref{vacuumspacetime}), except that the tensor indices read ``$AB$'' instead of
``$\alpha\beta$.''  The new question then arises, as to the nature of any possible
higher-dimensional matter fields $T_{AB}$.  Given that such fields are unobservable
in principle (to us as 4D observers), the most economical approach is to assume that
they do not exist.  This approach allows us to interpret matter and energy in the
4D world as manifestations of pure geometry in an empty higher-dimensional one.
Setting both $T_{AB}$ and $\Lambda$ to zero, we obtain
\bea
R_{AB} = 0.
\label{rabzero}
\eea
Contained in this elegant expression for 5D are fifteen relations which serve in principle
to determine the fifteen independent potentials $g_{AB}$.

Interest in higher-dimensional theories of gravity has spiked upward in recent years,
following the relaxation of the cylinder condition and its associated mechanism of
compactification.  Along with this has come the realization that mass may be geometrizable
in the same sense that time was geometrized by Minkowski (Wesson 2002a).  The extra coordinate
associated with mass is not necessarily lengthlike, but may be rewritten as a length coordinate
by means of appropriate dimension-transposing parameters.  Thus, $\ell$ is associated with
the rest mass $m$ of a classical particle via $\ell_{E}=Gm/c^2$ or $\ell_{P}=h/mc$,
which are of particular interest and are commonly referred to as the Einstein and
Planck gauges, respectively.  The former embeds the Schwarzschild solution, recognizes
the weak equivalence principle as a symmetry of the 5D metric, and gives back 4D
geodesic motion.  The latter gauge leads to the quantization rule $\int mcds=nh$
where $n$ is dimensionless but may be rational, as well as the Heisenberg-type
relation $|dp_{\alpha}dx^{\alpha}|=h(dn)^{2}/cn$.  For completeness, we mention that $\ell$
can also be related to $m$ via the particle's five-momentum, modulo the invariance of the
square of the four-momentum (Ponce de Leon 2003a; Ponce de Leon 2003b).  This provides an
invariant definition of the particle's rest mass in 4D, and a basis for calculating its
variation along the particle's worldline.  This mass variation is analogous to the removal
of absolute time that followed Minkowski's introduction of spacetime, and the new approach
is often called Space-Time-Matter (STM) theory for that reason.  In STM theory, the world
presumably appears four-dimensional because the effects of the fifth dimension are
suppressed by tiny factors of $G\dot{m}/c^3$ or $h\dot{m}/m^2c^2$, just as relativistic
effects in 4D Minkowski spacetime only appear at high velocities because they are suppressed
by factors of $\dot{x}/c$.  We are thus following a traditional line, but extended by (at least)
one extra dimension.

The mathematical principle underlying STM theory, known as Campbell's theorem
(Rippl et al. 1995; Romero et al. 1996; Lidsey et al. 1997; Seahra and
Wesson 2003a), states that any analytic $(N-1)$D Riemannian manifold can be locally
embedded in an $N$D Riemannian manifold which is Ricci flat (i.e. $R_{AB}=0$).  In the
context of 5D STM theory, this means that a curved 4D manifold with sources can be
embedded in a 5D Ricci-flat one which is empty.  Mathematically, this involves the
reduction of the 5D vacuum field equations~(\ref{rabzero}), wherein solutions to the
4D field equations with matter are recovered.  We use up four of the available five
coordinate degrees of freedom to set the electromagnetic potentials to zero
($g_{4\alpha}=0$); a generalization to metrics with $g_{4\alpha}\neq0$ has been given
by Ponce de Leon (2002).  Here, the 5D line element can be expressed in terms of a
scalar field $\Phi$ in the form
\bea
dS^{2} = g_{\alpha\beta}(x^{\gamma},\ell)dx^{\alpha}dx^{\beta} 
       + \varepsilon\Phi^{2}(x^{\gamma},\ell)d\ell^{2},
\label{generalmetric}
\eea
where $\varepsilon^{2}=1$ is employed to allow for either signature on the fifth dimension.
Putting this 5D metric into the 5D vacuum field equations~(\ref{rabzero}), we recover
the 4D field equations~(\ref{geinstein}), with matter sources described by
\bea
\frac{8\pi G}{c^{4}}T_{\alpha\beta}
&=& \frac{\nabla_{\beta}(\partial_{\alpha}\Phi)}{\Phi}
- \frac{\varepsilon}{2\Phi^{2}}\Biggl\lbrace\frac{\stackrel{*}{\Phi}
    \stackrel{*}{g}_{\alpha\beta}}{\Phi}
- \stackrel{**}{g}_{\alpha\beta}
+ g^{\lambda\mu}\stackrel{*}{g}_{\alpha\lambda}\stackrel{*}{g}_{\beta\mu}\nonumber\\
&\phantom{=}& - \frac{1}{2}g^{\mu\nu}\stackrel{*}{g}_{\mu\nu}\stackrel{*}{g}_{\alpha\beta}
+ \frac{1}{4}g_{\alpha\beta}\left[\stackrel{*}{g}^{\mu\nu}\stackrel{*}{g}_{\mu\nu}
+ \left(g^{\mu\nu}\stackrel{*}{g}_{\mu\nu}\right)^{2}\right]\Biggr\rbrace\nonumber\\
\nabla_{\alpha}\nabla^{\alpha}\Phi
&=& -\frac{\epsilon}{2\Phi}\left[\frac{\stackrel{*}{g}^{\lambda\beta}\stackrel{*}{g}_{\lambda\beta}}{2}
    + g^{\lambda\beta}\stackrel{**}{g}_{\lambda\beta}
    - \frac{\stackrel{*}{\Phi}g^{\lambda\beta}\stackrel{*}{g}_{\lambda\beta}}{\Phi}\right].
\label{effect}
\eea
Provided this stress-energy tensor is used, Einstein's 4D field equations~(\ref{geinstein})
are automatically contained in the 5D vacuum equations (\ref{rabzero}).  The second equation
is a field equation that $\Phi$ satisfies, which comes from the component $R_{44}=0$.

The tensor $T_{\alpha\beta}$ has good properties and has, for instance, been shown to
satisfy requirements such as the first law of thermodynamics and Newton's second law
(Wesson 1992).  Many exact solutions of (\ref{rabzero}) are known, which have been
applied to systems ranging from cosmological fluids to elementary particles (Wesson
and Liu 1998).  The Ponce de Leon class of solutions have been particularly useful
in studying a wide range of cosmologies containing fluids with an isothermal equation
of state (Ponce de Leon 1988).  The Bianchi models, which could describe 3D homogeneous
nonisotropic cosmologies, have also been extended to 5D (Seiler and Roque 1991; Halpern
2001).  In addition, this approach has been applied to clusters of galaxies (Billyard and
Wesson 1996a) as well as the solar system (Kalligas et al. 1995; Liu and Mashhoon 2000).
In particular, the classical tests of general relativity (i.e. light deflection, time delay,
and perihelion shift) along with the geodetic precession test have recently been investigated
using the simplest 5D analogue of the 4D Schwarzschild metric (Liu and Overduin 2000).
This work has shown that STM theory is consistent with observation on astrophysical scales.

The dynamics of test particles on 5D manifolds and 4D submanifolds have also been
extensively studied, and two related results have emerged, as follows:  (a)~Reduction of the
5D equations of motion introduces a fifth force which acts parallel to the four-velocities
of the particles (Wesson et al. 1999; Mashhoon and Liu 2000), but
for certain parametrizations of the metric this
force can be made to disappear (Ponce de Leon 2001a; Seahra 2002).  (b)~When a manifold
is extended
from 4D to 5D the spacetime line element $ds$ is embedded in a larger line element $dS$,
and particles which are massive and move on timelike paths in 4D with $ds^{2}>0$ can
move on null paths in 5D with $dS^{2}=0$ (Billyard and Sajko 2001; Seahra and Wesson 2001;
Youm 2001).  This is generally the case provided that the fifth force acts and 4D paths are
allowed with $m=m(s)$.  That is, the mass of a particle is now a function of its position in
the manifold.  This is a realization of Mach's principle.  If the fifth dimension is physically
real, and matter really is a manifestation of the geometry of the higher-dimensional world,
then the local properties of a particle must depend on the global properties (distribution)
of all the matter in the universe.  We may express this mathematically by saying that
(unrestricted) 5D Riemannian geometry is rich enough algebraically to unify gravitation
and electromagnetism with their sources of mass and charge.

\section{Solitons and Dark Matter}

To model astrophysical phenomena such as the Sun or a black hole in Kaluza-Klein theory,
the spherically-symmetric Schwarzschild solution of general relativity must be extended
to higher dimensions.  In 4D, Birkhoff's theorem guarantees that the Schwarzschild metric
is both static and unique to within its single free parameter (the mass of the central
object).  This theorem does not hold in higher dimensions, however.  Thus,  solutions which
are spherically symmetric, in general depend on other parameters such as electric and scalar
charge (Gibbons 1982; Gross and Perry 1983; Matos 1987), and can also be time-dependent as
well as non-singular.

If the universe has more than four dimensions, then an object such as a black hole must
be modelled with a higher-dimensional analogue of the Schwarzschild metric.  Various
possibilities have been explored over the years, with most attention focusing on a
5D solution now generally known as the soliton metric
(Gross and Perry 1983; Sorkin 1983; Davidson and Owen 1985).  In isotropic form
this is given by
\bea
ds^{2} &=& \left(\frac{ar-1}{ar+1}\right)^{2\epsilon k}c^{2}dt^{2}
        - \left(\frac{a^{2}r^{2}-1}{a^{2}r^{2}}\right)^{2}
          \left(\frac{ar+1}{ar-1}\right)^{2\epsilon(k-1)}(dr^{2}+r^{2}d\Omega^{2})\nonumber\\
       &\phantom{=}& - \left(\frac{ar+1}{ar-1}\right)^{2\epsilon}d\ell^{2}.
\label{soliton}
\eea
Here there are three metric parameters $(a,\epsilon,k)$, of which only two
are independent because a consistency condition which follows from the field equations
requires that $\epsilon^{2}(k^{2}-k+1)=1$.  In the limit
where $\epsilon\rightarrow0$, $k\rightarrow\infty$ and $\epsilon k\rightarrow1$,
the solution (\ref{soliton}) reduces to the Schwarzschild solution on 4D hypersurfaces
$\ell=\mbox{constant}$.  In this limit the parameter $a$ can be identified as
$a=2c^{2}/GM_{S}$ where $M_{S}$ is the Schwarzschild mass. 

The Kaluza-Klein soliton differs from a conventional black hole in several key respects.
It contains a singularity at its center, but this center is located at $r=1/a$ rather than
$r=0$.  In fact, the point $r=0$ is not even part of the manifold, which ends at $r=1/a$.
The soliton's event horizon, insofar as it has one, also shrinks to a point at $r=1/a$.
For these reasons the soliton is better classified as a naked singularity than a black hole.
The most straightforward way to gain insight into the physical properties of these objects
is to study the behaviour of test particles in the surrounding spacetime.  All the classical
tests of general relativity, as well as geodetic precession, have now been analyzed for
this metric including the general situation in which the components of the test particle's
momentum and spin along the extra coordinate do not vanish.  Existing data on light-bending
around the Sun using very long-baseline radio interferometry (VLBI), ranging to Mars using
the Viking lander, and the perihelion precession of Mercury, all constrain the small parameter
$\epsilon$ associated with the extra part of the metric to values of $|\epsilon|\leq0.07$
for the Sun (Liu and Overduin 2000).  Advances in solar monitoring and VLBI
technology (Eubanks et al. 1997)
would improve this bound by a factor of $\sqrt{10}$.  And the upcoming launch of the Gravity
Probe~B satellite will allow measurements of $|\epsilon|$ for the Earth with an accuracy
of one part in $10^4$ or better.  Experimental limits on violations of the equivalence
principle by solar-system bodies (Overduin 2000), are capable of tightening these constraints
by three to six orders of magnitude.

Solitons have an extended matter distribution rather than having all their mass compressed
into the singularity, and this feature opens up a second avenue of investigation into their
properties as dark-matter candidates (Wesson 1994).  From the time-time component
of the induced stress-energy tensor~(\ref{effect}), the density of the induced matter
associated with the solution (\ref{soliton}) can be worked out as a function of radial distance
$r$.  The density is
\bea
\rho_{S}(r) = \frac{c^{2}\epsilon^{2}k a^{6}r^{4}}{2\pi G(ar-1)^{4}(ar+1)^{4}}
              \left(\frac{ar-1}{ar+1}\right)^{2\epsilon(k-1)}.
\label{solitonrho}
\eea
From the other components of $T_{\alpha\beta}$ it is found that this fluid is anisotropic
($T_{11}\neq T_{22}$), defining an average pressure
$p_{S}=(p_{1}+p_{2}+p_{3})/3=\rho_{S}c^{2}/3$. The matter associated with the soliton
therefore has a radiation-like equation of state.  (In this respect solitons more closely
resemble primordial black holes, which form during the radiation-dominated era, than
conventional ones which arise as the endpoint of stellar collapse.)  The components of
(\ref{effect}) can also be used to calculate the gravitational mass of the fluid inside
$r$ (Wesson 1999), which in this case is found to be
\bea
M_{g}(r) = \int\sqrt{g}d^{3}x(T_{0}^{0}-T_{i}^{i})
         = \frac{2c^{2}\epsilon k}{Ga}\left(\frac{ar-1}{ar+1}\right)^{\epsilon}.
\label{solitonM}
\eea
Here we are integrating over three-space with volume element $d^{3}x$, $g$ is the
absolute value of the determinant of the four-metric, and $i$ is a spatial index running
$1,\ldots,3$.  At large distances, the soliton density (\ref{solitonrho}) and gravitational
mass~(\ref{solitonM}) read
\bea
\rho_{S}(r) \rightarrow \frac{c^{2}\epsilon^{2}k}{2\pi Ga^{2}r^{4}}
\quad
\mbox{and}
\quad
M_{g}(r) \rightarrow M_{g}(\infty) = \frac{2c^{2}\epsilon k}{Ga}.
\label{solitonproperties}
\eea
The second of these expressions shows that the asymptotic value of $M_{g}$ is, in
general, not the same as $M_{S}$ [i.e. $M_{g}(\infty)=\epsilon k M_{S}$ for $r>>1/a$],
but reduces to it in the limit $\epsilon k\rightarrow1$.  Viewed in 4D, the soliton
resembles a hole in the geometry surrounded by a spherically-symmetric ball of
ultra-relativistic matter whose density falls off as $1/r^{4}$.  If the universe does
have more than four dimensions, then objects like this should be common, being generic
to 5D Kaluza-Klein gravity in the same way that black holes are to 4D general
relativity.  It is then natural to ask whether these objects could in fact compose the
dark matter that fills most of the universe.

One way to find out more about the properties of dark-matter solitons is to estimate
the size of their effects on the cosmic microwave background (CMB).
This can be done using conventional methods (Overduin and Wesson 2003), assuming
that the fluid making up the soliton is composed of photons (although
ultra-relativistic particles such as neutrinos could also be considered in principle).
In the absence of a detailed spectral model we proceed bolometrically.  Putting the
second of equations~(\ref{solitonproperties}) into the first gives
\bea
\rho_{S}(r) \simeq \frac{GM_{g}^{2}}{8\pi c^{2}k r^{4}}.
\label{bolometry}
\eea
Numbers can be attached to the quantities $k$, $r$, and $M_{g}$ as follows.
The first ($k$) is technically a free parameter, but a natural choice from
the physical point of view is $k\sim 1$.  For this case, the consistency
relation implies $\epsilon\sim 1$ as well, which guarantees that the asymptotic
gravitational mass is close to its Schwarzschild mass.  To obtain a value for $r$,
it is sufficient to assume that solitons are distributed homogeneously through
space with average separation $d$ and mean density
$\bar{\rho}_{S}=\Omega_{S}\rho_{crit,0}=M_{S}/d^{3}$.  Since $\rho_{S}$ drops as
$1/r^{4}$ whereas the number of solitons climbs only as $r^{3}$, the local density
is largely determined by the nearest one.  Therefore $r$ can be replaced by
$d=(M_{S}/\Omega_{S}\rho_{crit,0})^{1/3}$.  The last unknown in (\ref{bolometry})
is the soliton mass $M_{g}$ ($=M_{S}$ if $k=1$).  Theoretical work suggests that
solitons are likely to be associated with dark matter on scales larger than the
solar system (Liu and Overduin 2000; Overduin 2000), so a natural identification
is with the dark-matter mass of the galactic halo, $M_S\simeq 2\times10^{12}M_{\odot}$.
Assuming that solitons make up $\Omega_S\simeq 0.3$ (a consensus value for
the dark-matter density today), we arrive at
\bea
\rho_S/\rho_{CMB} \simeq 7 \times 10^{-6}.
\label{solitonCMB}
\eea
Here $\rho_{CMB}$ is the energy density in CMB photons and accompanying neutrinos.
The limit~(\ref{solitonCMB}) is essentially the same as the upper bound
set on anomalous contributions to the CMB by COBE and other experiments,
so we infer that solitons are probably not much more massive than galaxies.  Similar
arguments can be made on the basis of tidal effects and gravitational lensing
(Wesson 1994).  To go further and put more detailed constraints on these candidates
from background radiation or other considerations will require more
investigation of their microphysical properties.

Supersymmetric particles such as gravitinos and neutralinos, if they exist, could also
provide the dark matter necessary to explain the dynamics of galaxies.  This kind of
dark matter may not be so dark, because the particles in question are unstable to
decay in non-minimal supersymmetric theories, and will contribute to the intergalactic
radiation field.  Observations of the latter can be used to constrain supersymmetric
weakly-interacting particles (WIMPs), with the result that gravitinos and neutralinos
remain viable as dark-matter candidates only if they decay on timescales of order
$10^{11}$~years or longer (Overduin and Wesson 2003).  Other leading dark-matter
candidates such as massive neutrinos, axions, primordial black holes and vacuum energy
encounter serious problems, and are constrained to narrow ranges in parameter space, or
ruled out altogether by data on the extragalactic background light and the CMB.  In this
sense, higher-dimensional candidates based on supersymmetry and STM theory are favoured
over traditional 4D ones.  While there are other candidates, the identification of dark
matter is an important experimental probe of gravitational theories in higher dimensions.

\section{Particle Dynamics}

The motion of test bodies is governed by the geodesic equation, which can be derived
from a least-action principle $\delta[\smallint\!m\,ds]=0$.  Here $m$ is the particle mass
(usually assumed constant) and the proper distance $s$ from A to B may be expressed in terms
of the Lagrangian density by $s=\smallint_A^B{\cal L}(\dot{x}^{\alpha},x^{\alpha})d\lambda$.
As usual, $\lambda$ is an affine parameter (i.e. one related to proper time $\tau$ via
$\lambda=a\tau +b$) and an overdot signifies $d/d\lambda$.
In 4D general relativity this prescription leads to
\bea
\frac{d^{2}x^{\alpha}}{d\lambda^{2}} + \Gamma_{\beta\rho}^{\alpha}\frac{dx^{\beta}}{d\lambda}
\frac{dx^{\rho}}{d\lambda} = 0,
\label{geo}
\eea
where $\Gamma_{\beta\rho}^{\alpha}$ are the Christoffel symbols of the second kind.
This may be seen as a curved-spacetime analogue of Newton's second law $F=ma$ for
gravitational forces, with the term $d^{2}x^{\alpha}/d\lambda^{2}$ identified with the
four-acceleration of the particle, and the Christoffel coefficients playing the role of
generalized ``gravitational forces''.  (Note, however, that neither ordinary derivatives
nor Christoffel symbols are tensors.)
Now equation~(\ref{geo}) is independent of $m$, which implies that all test bodies,
regardless of their mass, travel along the same path in the same gravitational field.
This is the same as what Galileo found over four hundred years ago, but we should
note that it follows here only because $m$ has been taken as constant in the original
variational principle.

Before extending the dynamics of test particles from 4D to 5D, we introduce a metric
which will simplify the analysis without loss of generality.  This is commonly known
as the canonical metric and reads
\bea
dS^{2} = \frac{\ell^{2}}{L^{2}}g_{\alpha\beta}(x^{\alpha},\ell)dx^{\alpha}dx^{\beta} - d\ell^{2},
\label{canonical}
\eea
where $L$ is a length that has been introduced for dimensional consistency.
All five available coordinate degrees of freedom have been used to set $g_{4\alpha}=0$
and $g_{44}=-1$.  Physically this suppresses potentials of electromagnetic type and
flattens the potential of scalar type.  Equation~(\ref{canonical}) has the virtue of
greatly simplifying the algebra associated with the field equations (\ref{rabzero}).
For a certain class of cosmological solutions the length $L$ is identified with the
cosmological constant via $L=\sqrt{3/\Lambda}$ (more on this shortly), while for other
solutions $L$ is identified as the characteristic size of the four-space.  Now, if $\ell$
is related to $m$ then the 4D part of the 5D interval is $(\ell/L)ds$, which defines a
momentum space rather than a coordinate space, thus describing particle dynamics in terms
of four-momenta rather than four-velocities.  It is also worth noting that the energy of a
particle moving with velocity $v$ in the Minkowski limit is $E=\ell/\sqrt{1-v^{2}}$ in 5D
(with $c=1$), which is the same as the 4D expression if $\ell\sim m$.

The equations of motion are obtained exactly as in 4D from a least-action principle
of the form $\delta[\smallint dS]=0$, where the Lagrangian density is now
\bea
\mathcal{L} = \frac{dS}{d\lambda}
= \sqrt{\frac{\ell^{2}}{L^{2}}g_{\alpha\beta}
\frac{dx^{\alpha}}{d\lambda}\frac{dx^{\beta}}{d\lambda} 
- \left(\frac{d\ell}{d\lambda}\right)^{2}}.
\eea
Here $\lambda$ is an affine parameter along the path as before.  In general,
$\mathcal{L}\equiv\mathcal{L}(\dot{x}^{A},x^{A})$ where $\dot{x}^{A}=dx^{A}/d\lambda$
is the particle's five-velocity.  The action will be an extremum if $\mathcal{L}$
satisfies the 5D Euler-Lagrange equations
\bea
\frac{\partial\mathcal{L}}{\partial x^{A}}
- \frac{d}{d\lambda}\left(\frac{\partial\mathcal{L}}{\partial\dot{x}^{A}}\right) = 0.
\eea
One choice of parameter is $\lambda=S$, and this yields the 5D geodesic equation
\bea
\frac{d^{2}x^{A}}{dS^{2}} + \Gamma_{BC}^{A}\frac{dx^{B}}{dS}\frac{dx^{C}}{dS} = 0,
\eea
where 5D Christoffel coefficients are defined in exactly the same way as their 4D
counterparts.  On the other hand, if we wish to understand the dynamics of 4D particles
in terms of four-velocities $\dot{x}^{\alpha}=dx^{\alpha}/ds$, then a more transparent
choice of parameter is $\lambda=s$.  Using this framework with the canonical metric
(\ref{canonical}), the Euler-Lagrange equations yield two equations: one that involves
motion in 4D coordinates modified by the addition of a force-like quantity; and another
that involves the motion in the fifth coordinate.  These equations are given by
\bea
\frac{d^{2}x^{\alpha}}{ds^{2}}
+ \Gamma_{\beta\gamma}^{\alpha}\frac{dx^{\beta}}{ds}\frac{dx^{\gamma}}{ds}
&=& f^{\alpha}\nonumber\\
f^{\alpha}
&\equiv& \left(\frac{1}{2}\frac{dx^{\alpha}}{ds}\frac{dx^{\mu}}{ds} - g^{\alpha\mu}\right)
\frac{d\ell}{ds}\frac{dx^{\beta}}{ds}\frac{\partial g_{\mu\beta}}{\partial\ell}\nonumber\\
\frac{d^{2}\ell}{ds^{2}} - \frac{2}{\ell}\left(\frac{d\ell}{ds}\right)^{2} + \frac{\ell}{L^{2}}
&=& -\frac{1}{2}\left[\frac{\ell^{2}}{L^{2}} - \left(\frac{d\ell}{ds}\right)^{2}\right]
\frac{dx^{\alpha}}{ds}\frac{dx^{\beta}}{ds}\frac{\partial g_{\alpha\beta}}{\partial\ell}.
\label{reduction}
\eea
An inspection of the force-like term $f^{\alpha}$ reveals that there is a departure from
ordinary 4D geodesic motion if (a) the 4D metric depends on the fifth coordinate
($\stackrel{*}{g}_{\alpha\beta}\neq0$), and (b) there is motion in the fifth dimension
($d\ell/ds\neq0$).  A solution of the second equation would then give the rate of motion in
the fifth dimension.  It should be noted, however, that the fifth force is different from
others in 4D dynamics.  Normally a non-gravitational force acting on a particle in 4D is
equivalent to the four-acceleration through the relation
$F^{\alpha}=Du^{\alpha}/ds\equiv u^{\beta}\nabla_{\beta}u^{\alpha}$ (Wald 1984).  However,
the four-velocity is normalized such that $u_{\alpha}u^{\alpha}=1$ and so
$(\nabla_{\beta}u^{\alpha})u_{\alpha}=0$ and it follows that $F^{\alpha}u_{\alpha}=0$.  This
means that the force is orthogonal to the four-velocity.  The corresponding condition in 5D
is $F^{A}u_{A}$, and with $F^{A}=(f^{\alpha},f^{4})$, $u_{A}=(u_{\alpha},u_{4})$ it follows
that $f^{\alpha}u_{\alpha}=-f^{4}u_{4}\neq0$.  This shows that the force $f^{\alpha}$ in
(\ref{reduction}) does not act orthogonally to the four-velocity.  In fact, this force acts
parallel to the four-velocity $u^{\alpha}$ for metrics in canonical form, so it is natural to
express its effects in terms of the momentum or (inertial) rest mass of the particle that feels
it (Wesson 1999; Youm 2000).  This is the reason why $m$ can be related to $\ell$ or its rate
of change, depending on the coordinates.  Furthermore, it has been shown that $f^{\alpha}$
is in general not equivalent to the four-acceleration in 5D, and as a result does not satisfy
the standard tensor transformation rule if the 4D coordinate transformation depends on $\ell$.
As a result, $f^{\alpha}$ is not a four-vector (Seahra 2002).  It is important to note that
the fifth force does not manifest itself for a certain class of metrics based on pure canonical
coordinates (Mashhoon et al. 1994), for metrics where the coordinates can be chosen so as to
make the velocity in the extra dimension comoving, or for metrics parametrized in such a way
as to make it disappear (Ponce de Leon 2001a).  In general, however, the fifth force does exist
for 5D metrics which depend on the extra coordinate $\ell$ (Billyard and Sajko 2001; Wesson 2002b)
as well as brane-world theory (Maartens 2000; Belayev 2001; Chamblin 2001).  That is, astrophysics
in 5D is modified by the presence of an extra force.

As already mentioned, the embedding of spacetime in a Ricci-flat 5D manifold guarantees
that the line element $dS$ will contain $ds$.  Accordingly, particles moving on null paths
in 5D $(dS^{2}=0)$ will appear as massive particles moving on timelike paths in 4D 
$(ds^{2}>0)$, provided that $\ell$ is spacelike and that the particles have $P_{4}\neq0$ (see
below).  Otherwise, the particles will appear as massless particles moving on null paths in 4D
$(ds^{2}=0)$.  Consequently, massive particles appear in 4D with $m=m(s)$.  This suggests an
intriguing possibility, that the presence of the fifth dimension could in principle be detected
as a variation of the (rest) mass of a particle with proper time.  The basic idea is that an
anomalous force due to the fifth dimension could show up as a timelike component proportional
to the particle's four-velocity with a constant of proportionality $m^{-1}dm/ds$.  This would
correspond to a slow variation of the rest mass of a particle over long timescales, as
interpreted in four-dimensional terms.  It should be noted that the condition $dS^{2}=0$ is
compatible with the conventional 4D relations for mass, energy and momentum, but also means that
the particle's mass is zero in 5D.  This is to be expected, as the embedding theorem requires
that the 5D manifold must be Ricci-flat and empty.  More recent work on dynamics in the
braneworld and other 5D models (Seahra 2002) reveals that in addition to mass variations, there
may also appear variations in the spin of a point-like gyroscope, which take place over long
timescales.  This can be measured by projects such as Gravity Probe B, where the dynamics of a
gyroscope will be measured from aboard a satellite in Earth orbit.  Alternatively, the gradual
change of angular momentum predicted by 5D theory could be investigated via observations of
spiral galaxies at high redshifts.

\section{The Cosmological Constant}

The cosmological constant $\Lambda$ on the right-hand side of equation~(\ref{geinstein2})
determines the energy density and pressure of the vacuum via
\bea
\rho_{v} = -p_{v} = \frac{\Lambda c^{2}}{8\pi G}.
\eea
This description fits naturally into 4D physics insofar as the vacuum is treated as a
``classical'' entity where gravity is associated with all forms of energy.  But quantum
field theory involves non-gravitational fields that have energy even at the absolute zero
of temperature.  For example, quantum electrodynamics associates an energy $\hbar\omega/2$
with each mode of frequency $\omega$ at absolute zero, and this multiplied by the density
of modes gives a highly energetic zero-point field.  The other interactions are presumed to
have similar zero-point fields.  The cosmological-constant problem is thus: astrophysical
data shows $|\Lambda|$ (and hence the density of the vacuum) to be small, whereas quantum
field theories predict a massive value for the zero-point fields. Various resolutions to
this problem have been proposed.  For example, quantum processes with their appropriate
expectation values might effectively force the effective value of $\Lambda$ toward zero with
time.  This is theoretically possible, perhaps in a space with a changeable topology.
But it would imply that we live at a special time, since the magnitude-redshift relation for
Type~Ia supernovae and other astrophysical data demonstrate that $\Lambda$ (while tiny
in comparison to theoretical expectations) is nevertheless nonzero.

A more recent approach to the problem involves the reduction of a non-compact $N$D field
theory (with $N\geq5$) to 4D, which can yield an effective 4D $\Lambda$ that is small
(Wesson and Liu 2001).
The mismatch between energy densities derived from quantum field theory and general
relativity may then be a consequence of restricting the physics to 4D.  Let us now proceed
to see how the field equations (\ref{rabzero}) can be reduced to 4D in such a way that the
cosmological constant is identified with the cosmological length scale, or equivalently the
coordinates of a particle (modulo a conformal factor).  Using the canonical coordinates
defined in (\ref{canonical}), we take
$g_{\alpha\beta}=(\ell^{2}/L^{2})\tilde{g}_{\alpha\beta}(x^{\alpha},\ell)$ and $g_{44}=-1$.
Imposing the condition $\stackrel{*}{\tilde{g}}_{\alpha\beta}=0$ it can be shown that
equations~(\ref{vacuumspacetime}) are satisfied with
\bea
\tilde{\Lambda} = \frac{3}{L^{2}}.
\label{lambda1}
\eea
If $L$ is large then $\tilde{\Lambda}$ is small as required by observational data.
However, $\tilde{\Lambda}$ cannot be zero if we require that the 4D part of the 5D metric
be finite in the solar system (Liu and Mashhoon 2000).  For in this case the field equations
$R_{AB}=0$ are satisfied with $\stackrel{*}{\tilde{g}}_{\alpha\beta}=0$ and (\ref{lambda1})
by the solution
\bea
dS^{2} &=& \frac{\tilde{\Lambda}\ell^{2}}{3}
         \left\lbrace\left[1-\frac{2M}{r}-\frac{\tilde{\Lambda}r^{2}}{3}\right]dt^{2}
                   - \left[1-\frac{2M}{r}-\frac{\tilde{\Lambda}r^{2}}{3}\right]^{-1}dr^{2}
                   - r^{2}d\Omega^{2}\right\rbrace\nonumber\\
&\phantom{=}& - d\ell^{2}.
\label{sembedding}
\eea
Here $M$ is the mass, so this is a 5D embedding for the 4D Schwarzschild-de Sitter
solution.  Furthermore, $\stackrel{*}{\tilde{g}}_{\alpha\beta}=0$ so that the fifth force
of (\ref{reduction}) is zero.  Hence the 4D motion is described by the 4D geodesic equation
(\ref{geo}), and it follows that there is no way to tell by the classical tests of GR if the
solar system is described by the 5D metric (\ref{sembedding}) or only its 4D part.  Note that
the effective cosmological constant defined by (\ref{lambda1}) was obtained by using the pure
4D part $\tilde{g}_{\alpha\beta}(x^{\alpha})$ of the 5D metric (\ref{canonical}).  If we use
$g_{\alpha\beta}$ instead, then we find that $R_{\alpha\beta}=(3/\ell^{2})g_{\alpha\beta}$.
This describes a vacuum spacetime with a cosmological constant
\bea
\Lambda = \frac{3}{\ell^{2}}.
\label{lambda2}
\eea
Here $\Lambda$ is a function of the fifth coordinate $\ell$, and in the static limit
the correspondence between the energy of a test particle in 4D and 5D requires the
identification $\ell\sim m$ where $m$ is the rest mass (Liu and Mashhoon 2000).
This
suggests that each particle of mass $m$ may determine its own value of $\Lambda$.

The signature of the 5D metric in STM theory has important implications for
the sign of the cosmological constant.  In older work, the signature was often taken
to be $(+----)$.  However, in modern work it is left general via $(+---\varepsilon)$.
In fact, there are well-behaved solutions with good physical properties which
describe galaxies in clusters or waves in vacuum
(Billyard and Wesson 1996b) 
which have signature $(+---+)$.  If we proceed as above with
$\stackrel{*}{\tilde{g}}_{\alpha\beta}=0$, but this time taking $g_{44}=\varepsilon$,
then the field equations~(\ref{vacuumspacetime}) define a cosmological constant
\bea
\tilde{\Lambda} = -\frac{3\varepsilon}{L^{2}}.
\label{cosmocon}
\eea
A measurement of the sign of the cosmological constant may thus also be a determination
of the signature of the (higher-dimensional) world.  Lower limits on the age of the
universe, as well as recent observational data on high-redshift supernovae, favour a
positive $\Lambda$.  This in conjunction with
equation~(\ref{cosmocon})
is consistent with a spacelike fifth dimension.

\section{Aspects of Quantum Theory}

The determination of the cosmological constant also has important consequences for
quantum physics, and we discuss this here along with some more general quantum aspects
of higher-dimensional gravity.  It is convenient to use only four of
the available five degrees of coordinate freedom, so that $g_{44}=g_{44}(x^{4})$ may
be related explicitly to the Higgs potential, as in newer versions of Kaluza-Klein
theory.  The problem remains mathematically general so long as 
$g_{\alpha\beta}=g_{\alpha\beta}(x^{\gamma},x^{4})$.  Exact solutions of the field
equations are known with this property, including cosmological ones which reduce
to Friedmann-Robertson-Walker (FRW) models on $x^{4}=\mbox{constant}$ hypersurfaces and
agree with observation (Wesson 1999).  However, the focus here is on particle dynamics;
and since observations indicate that there is no explicit incursion of the fifth dimension
into local spacetime, we can take $g_{\alpha\beta}=g_{\alpha\beta}(x^{\gamma})$, which
means that the weak equivalence principle is a symmetry of the 4D part of the 5D metric.
Without identifying $x^{4}=\ell$ explicitly at the outset, we expect that (a)~$x^{4}$
will be related to the (inertial) rest mass of a test particle, and (b)~$m=m(s)$.
The latter expression does not imply a violation of the usual condition
$g_{\alpha\beta}u^{\alpha}u^{\beta}=1$ (for four-velocities
$u^{\alpha}=dx^{\alpha}/ds$) because it is a normalization condition on the
velocities and not a coordinate condition on the metric.
Multiplying by $m^{2}$ gives $p^{\alpha}p_{\alpha}=m^{2}$ where
$p^{\alpha}=mu^{\alpha}$ with no restriction on $m=m(s)$.  In other words, the
energy $E^{2}=m^{2}c^{4}u^{0}u_{0}$ and three-momentum
$p^{2}=m^{2}c^{2}(u^{1}u_{1}+u^{2}u_{2}+u^{3}u_{3})$ can satisfy
$E^{2}-p^{2}c^{2}-m^{2}c^{4}=0$, even if the mass varies in spacetime.  This last relation
describes real particles and implies that our 5D metric should have signature
$(+----)$, thus making the fifth dimension spacelike, in agreement with the
conclusion obtained from the sign of $\Lambda$ in the previous section.

Based on the preceding, and introducing a constant $L$ for dimensional consistency, a
line element can be considered with the general form
\bea
dS^{2} = \frac{L^{2}}{\ell^{2}}g^{\alpha\beta}(x^{\gamma})dx^{\alpha}dx^{\beta}
       - \frac{L^{4}}{\ell^{4}}d\ell^{2}.
\label{planckianmetric}
\eea
The generalized five-momenta are given by
\bea
P_{\alpha} &=& \frac{\partial\mathcal{L}}{\partial(dx^{\alpha}/ds)}
           = \frac{2L^{2}}{\ell^{2}}g_{\alpha\beta}\frac{dx^{\beta}}{ds}\nonumber\\
P_{4} &=& \frac{\partial\mathcal{L}}{\partial(d\ell /ds)}
         = -\frac{2L^{4}}{\ell^{4}}\frac{d\ell}{ds}.
\label{fivemomenta}
\eea
These are associated with the Lagrangian density $\mathcal{L}=(dS/ds)^{2}$ and
define a 5D scalar analogous to the one used in 4D quantum mechanics:
\bea
\int P_{A}dx^{A} = \int(P_{\alpha}dx^{\alpha}+P_{4}d\ell)
                 = \int\frac{2L^{2}}{\ell^{2}}
                   \left[1-\left(\frac{L}{\ell}\frac{d\ell}{ds}\right)^{2}\right]ds.
\label{fivescalar}
\eea
This is zero for $dS^{2}=0$ and hence equation~(\ref{planckianmetric}) gives
\bea
\ell = \ell_{0}\mbox{e}^{\pm s/L}
\quad
\rightarrow
\quad
\frac{d\ell}{ds} = \pm\frac{\ell}{L},
\label{dlds}
\eea
where $\ell_{0}$ is a constant.  Thus, the mass variation of a test particle is slow
if $s/L\ll1$.  Now, the 4D quantity $\smallint p_{\alpha}dx^{\alpha}$ which corresponds to
(\ref{fivescalar}) is non-zero for a massive particle, and from (\ref{dlds}) is found to be
\bea
\int p_{\alpha}dx^{\alpha} = \int mu_{\alpha}dx^{\alpha} = \int\frac{hds}{c\ell}
                           = \pm\frac{h}{c}\frac{L}{\ell}.
\label{fourscalar}
\eea
The fact that this can be positive or negative goes back to (\ref{dlds}), but since the
motion is reversible it is convenient to suppress the sign for what follows.  The goal
is to have (\ref{fourscalar}) reproduce the usual 4D action of the form $mcds$, and this
is accomplished using the Compton wavelength $\ell=h/mc$ of the particle, which
has finite energy in 4D but zero energy in 5D because $\smallint P_{A}dx^{A}=0$.
Therefore equation~(\ref{fourscalar}) becomes
\bea
\int mcds = nh,
\label{action}
\eea
where we have introduced the number $n\equiv L/\ell$.  This says that the conventional
action of particle physics in 4D follows from a null line element (\ref{planckianmetric})
in 5D.

Coming this far, it is instructive to investigate the properties of the second scalar
quantity $dp_{\alpha}dx^{\alpha}$.  Following the same procedure as outlined above,
there comes
\bea
dp_{\alpha}dx^{\alpha} = \frac{h}{c}\left(\frac{du_{\alpha}}{ds}\frac{dx^{\alpha}}{ds}
                       - \frac{1}{\ell}\frac{d\ell}{ds}\right)\frac{ds^{2}}{\ell}.
\eea
Here the first term inside the parentheses is zero if the acceleration is zero, or if the
scalar product with the velocity is zero as in conventional 4D dynamics.  Nevertheless,
there is still a contribution from the second term inside the parentheses which is due
to the change in mass of the particle.  This anomalous contribution has magnitude
\bea
|dp_{\alpha}dx^{\alpha}| = \frac{h}{c}\Bigg|\frac{d\ell}{ds}\Bigg|
                                      \left(\frac{ds}{\ell}\right)^{2}
                                    = \frac{h}{c}\frac{ds^{2}}{L\ell}
                                    = n\frac{h}{c}\left(\frac{d\ell}{\ell}\right)^{2},
\label{heisenbergtype}
\eea
where (\ref{dlds}) and $n=L/\ell$ have been used.  The latter implies
$dn/n=-d\ell/\ell=dK_{4}/K_{4}$ where $K_{4}\equiv1/\ell$ is the
wavenumber for the extra dimension.  Equation~(\ref{heisenbergtype}) is a Heisenberg-type
relation, and can be written as
\bea
|dp_{\alpha}dx^{\alpha}| = \frac{h}{c}\frac{(dn)^{2}}{n}.
\label{heisenberg}
\eea
Looking back at the 5D line element (\ref{planckianmetric}), it is apparent that $L$ is
a length scale not only for the extra dimension but also for the 4D part of the manifold.
(There may be other scales associated with the sources for the potentials that are
contained in $g_{\alpha\beta}$ which may define a scale via the 4D Ricci scalar $R$, but
the 5D field equations are expected to relate $R$ to $L$.)  As the particle moves through
spacetime, it therefore ``feels'' $L$, and this is reflected in the behaviour of its mass
and momentum.  The relations (\ref{action}) and (\ref{heisenberg}), which arise from the
mass parametrization $\ell=h/mc$, quantify this statement and prompt the conjecture that
classical 5D physics may lead to quantized 4D physics.  For this reason we refer to
equation~(\ref{planckianmetric}) with the parametrization $\ell_{P}\equiv\ell=h/mc$
as the Planck gauge.

If the particle is viewed as a wave, its four-momenta are defined by the de Broglie
wavelengths and its mass is defined by the Compton wavelength.  The relation $dS^{2}=0$
for (\ref{planckianmetric}) is equivalent to $P_{A}P^{A}=0$ or $K_{A}K^{A}=0$.  The
question then arises whether the waves concerned are propagating in an open topology
or trapped in a closed topology.  In the former case, the wavelength is not constrained
by the geometry, and low-mass particles can have large Compton wavelengths with
$\ell>L$ and $n=L/\ell<1$.  In the latter case, the wavelength cannot exceed the
confining size of the geometry, and high-mass particles have small Compton wavelengths
with $\ell\leq L$ and $n\geq1$.  By (\ref{heisenberg}) the former case obeys the
conventional uncertainty principle while the latter case violates it.  This subject
clearly needs further study, but for now the former case is tentatively identified
as applying to real particles, and the latter case as applying to virtual particles.

The fundamental mode ($n=1$) deserves special attention.  This can be studied using
equations (\ref{action})-(\ref{heisenbergtype}).  In particular, the parametrization
$\ell=h/mc$ causes (\ref{heisenbergtype}) to become
\bea
|dp_{\alpha}dx^{\alpha}| = \frac{h}{c}\frac{ds^{2}}{L\ell} = m\frac{ds^{2}}{L}.
\eea
Dividing by $ds$ and identifying $u^{\alpha}=dx^{\alpha}/ds$ gives
\bea
|dp_{\alpha}u^{\alpha}| = m\frac{ds}{L}.
\eea
With $dp_{\alpha}=dmu_{\alpha}+mdu_{\alpha}$ and $u_{\alpha}u^{\alpha}=1$ we get
$|dm|=mds/L$, and with (\ref{action}) we finally have
\bea
|m|=(\smallint mcds)/cL=nh/cL.
\eea
For $n=1$ this defines a fundamental unit of mass $m_{0}=h/cL$.  In general $L$ is a scale
set by the problem, in analogy with the ``box'' size in old wave mechanics.  In cosmology,
however, $L$ is related to the cosmological constant.  As with the canonical metric
(\ref{canonical}) in the previous section, the metric (\ref{planckianmetric}) identifies
$\Lambda=3/L^{2}$.  The cosmological value of $L=\sqrt{3/\Lambda}$ is in fact a maximum for
this parameter, thus defining a minimum for $m_{0}$ which applies even to particle physics.
Astrophysical data indicate a positive value for $\Lambda$ of approximately $3\times10^{-56}$
cm$^{-2}$, but in view of observational uncertainties this should be taken as a constraint
rather than a determination (Lineweaver 1998; Chiba and Yoshii 1999; Overduin 1999; Eppley
and Partridge 2000).  The noted value corresponds to a vacuum density of
$\Lambda c^{2}/8\pi G\simeq 2\times10^{-29}$ gcm$^{-3}$, close to that required for closure.
The unit mass involved is therefore
\bea
m_{0} = \frac{h}{cL} = \frac{h}{c}\sqrt{\frac{3}{\Lambda}} \simeq 2\times10^{-65}
\quad
\mbox{g}.
\label{quantummass}
\eea
This is too small to be detected using current techniques, and explains why mass does
not appear to be quantized.

The mass unit (\ref{quantummass}) of order $10^{-65}$ g follows from the length scale
$\Lambda^{-1/2}$ of order $10^{28}$~cm, or equivalently the timescale of order
$10^{18}$~s which is the age of the universe.  However, a more detailed analysis
might alter these numbers somewhat.  For example, a more detailed analysis might involve
the size of the particle horizon at the current epoch or the current size of the
cosmological constant in models where this parameter varies
(Overduin and Cooperstock 1998; Overduin 1999).
But equation~(\ref{quantummass}) is based on astrophysical data and may be expected
to hold to within an order of magnitude.  It should be noted that the interaction
between a particle with a mass of order $m_{0}$ and a vacuum energy with density of order
$\Lambda c^{4}/8\pi G$, involves physics that is poorly understood.  Preliminary
discussions of this and related topics have been given recently
(Mansouri 2002).
The view suggested by STM theory is that a ``particle'' is just a localized concentration
of energy in a medium where the distinction between ordinary matter and vacuum is convenient
but artificial.  This comes from a comparison of equations~(\ref{canonical}) and
(\ref{planckianmetric}) with a more general line element of the form (\ref{generalmetric}).
Energy, defined as the quantity that curves 4D spacetime, consists in general of both
types of contributions.
After all, energy is a 4D concept that can be derived from 5D geometry, with Campbell's
theorem providing the link
(Campbell 1926; Rippl et al. 1995; Romero et al. 1996; Lidsey et al. 1997).  The small mass
(\ref{quantummass}) then simply reflects the small 4D curvature of the universe.
In other problems the parameter $L$ can, in general, be interpreted differently.
Consider for example the hydrogen atom, with a length scale of order $10^{-8}$~cm, and
a corresponding timescale of order $10^{-18}$~s.  The corresponding unit mass would differ
by many orders of magnitude from that given by $m_0$ above.  Nevertheless,
equation~(\ref{quantummass}) defines what could be considered an irreducible unit mass
set by the energy density of the universe, as measured by the cosmological constant.

An interesting aspect of STM theory is that any line element on a 5D manifold can be
brought into canonical form via appropriate coordinate transformations.  Transforming
the line element (\ref{planckianmetric}) via $\ell\rightarrow L^{2}/\ell$ then gives
the line element (\ref{canonical}).  It has been shown above that this metric also leads to
the identification $\Lambda=3/L^{2}$.  Furthermore, since Campbell's theorem states
that any solution of the 4D vacuum field equations can be embedded in a solution of the
5D vacuum field equations, this holds for the Schwarzschild solution as well (Seahra 2002).
Equation~(\ref{canonical}) is therefore relevant to gravitational problems; and with the
appropriate parametrization $\ell_{E}\equiv\ell=Gm/c^{2}$, is hence referred to
as the Einstein gauge.  In the general case where
$g_{\alpha\beta}=g_{\alpha\beta}(x^{\gamma},\ell)$ and $dS^{2}\neq0$, one expects
violations of the weak equivalence principle because the four-accelerations of test
particles in general depend on $\ell$ and may not be the same for all of them.  This is
borne out by the equations of motion (\ref{reduction}), which follow from the
reduction of the geodesic equation from 5D to 4D.  The two gauges (\ref{canonical}) and
(\ref{planckianmetric}) are related by the simple coordinate transformation
$\ell\rightarrow L^{2}/\ell$, and are therefore equivalent mathematically.  Physically,
however, this corresponds to changing the way the rest mass of a particle is described.
In old 4D theories of scalar-tensor and scale-invariant types, the mass
parametrizations
\bea
\ell_{E} = \frac{Gm}{c^{2}}
\quad
\mbox{and}
\quad
\ell_{P} = \frac{h}{mc}
\label{gauges}
\eea
were sometimes referred to as reflecting the use of gravitational and atomic units
(Hoyle and Narlikar 1974).  In the present approach, however, they refer to the use
of coordinates in an underlying theory which is 5D covariant.  Here, the Planck mass
defined by $m=\sqrt{hc/G}$ lacks physical meaning.  This quantity can be formed by taking
the ratio $\ell_{E}/\ell_{P}=Gm^{2}/hc$ and setting it equal to unity, but this involves
mixing coordinate systems, and is therefore badly defined.  In other words, while the
Einstein and Planck gauges (\ref{gauges}) are unique and well defined, a mixture of the
two that produces the Planck mass is ill-defined, suggesting that if this parameter has
physical meaning it does so in the context of an $N(>5)$D field theory.

\section{The Big Bang Revisited}
In preceeding sections, we have examined the main changes to astrophysics and particle
physics that follow from an extension of relativity from 4D to 5D, as used in STM and
membrane theories.  However, it is clear that the results we have reported
need to be put into the context of a global cosmology.  This subject requires exact
solutions of the 5D field equations (\ref{rabzero}), which should consistently embed known
solutions of the 4D field equations (\ref{geinstein}) so that we can recover agreement
with the large amount of precision data which is now available from observations.  Since
this subject is fairly technical and in a state of rapid development, we content ourselves
with an account of the main result: a new 5D view of the 4D big bang.

Embeddings of 4D solutions in 5D manifolds have been studied for many years.  Campbell
outlined the proof of an important theorem in 1926, which basically says that any solution
of the 4D field equations with matter (\ref{geinstein}) can be locally embedded in a
solution of the apparently empty equations (\ref{rabzero}).  Magaard gave an explicit and
rigorous proof of this theorem later, but the Campbell-Magaard theorem was unknown to
other workers when they inaugurated STM theory in 1992.  Tavakol and coworkers
drew attention to the theorem in a series of papers that started in 1995, and pointed out
that it not only provides a way to go from 5D to 4D, but also in principle to go from 4D
to lower-dimensional gravity.  (Such models are widely regarded as simplistic by those who
do research, but may have value to those who do pedagogy.)  Ponce de Leon in 1988 had in
the meantime found a class of 5D cosmological models which reduce to the standard 4D
FRW ones on hypersurfaces where the extra coordinate is held fixed
(i.e. on those surfaces defined by $\ell=\ell_{0}=\mbox{constant}$).  The Ponce de Leon
solutions were used to study the big bang by Seahra and Wesson (2002).
They showed the result that curved 4D FRW models with a big bang can be embedded
in a flat 5D model without one.

\begin{figure}
\centering
\fbox{\includegraphics[width=2.2in]{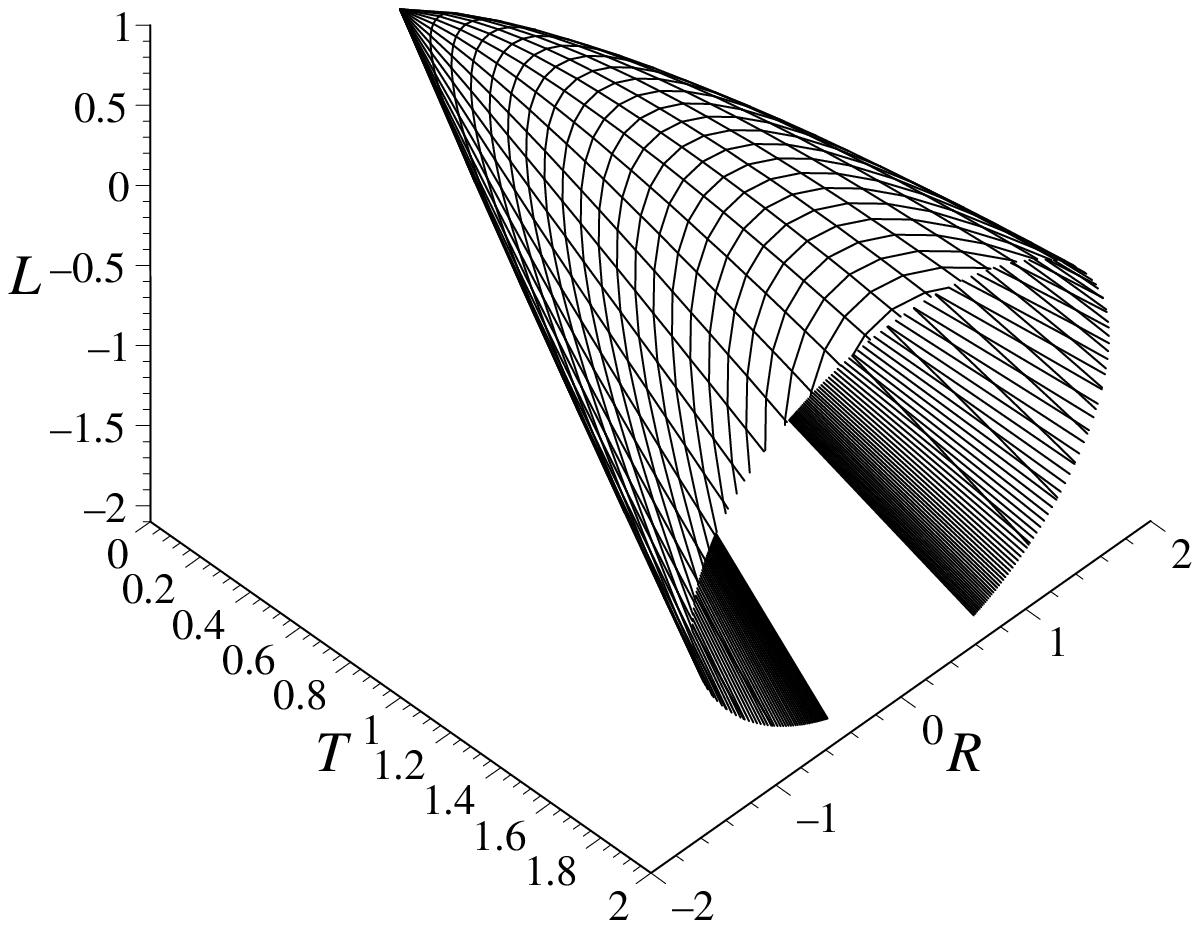}}
\hspace{0.3in}
\fbox{\includegraphics[width=2.2in]{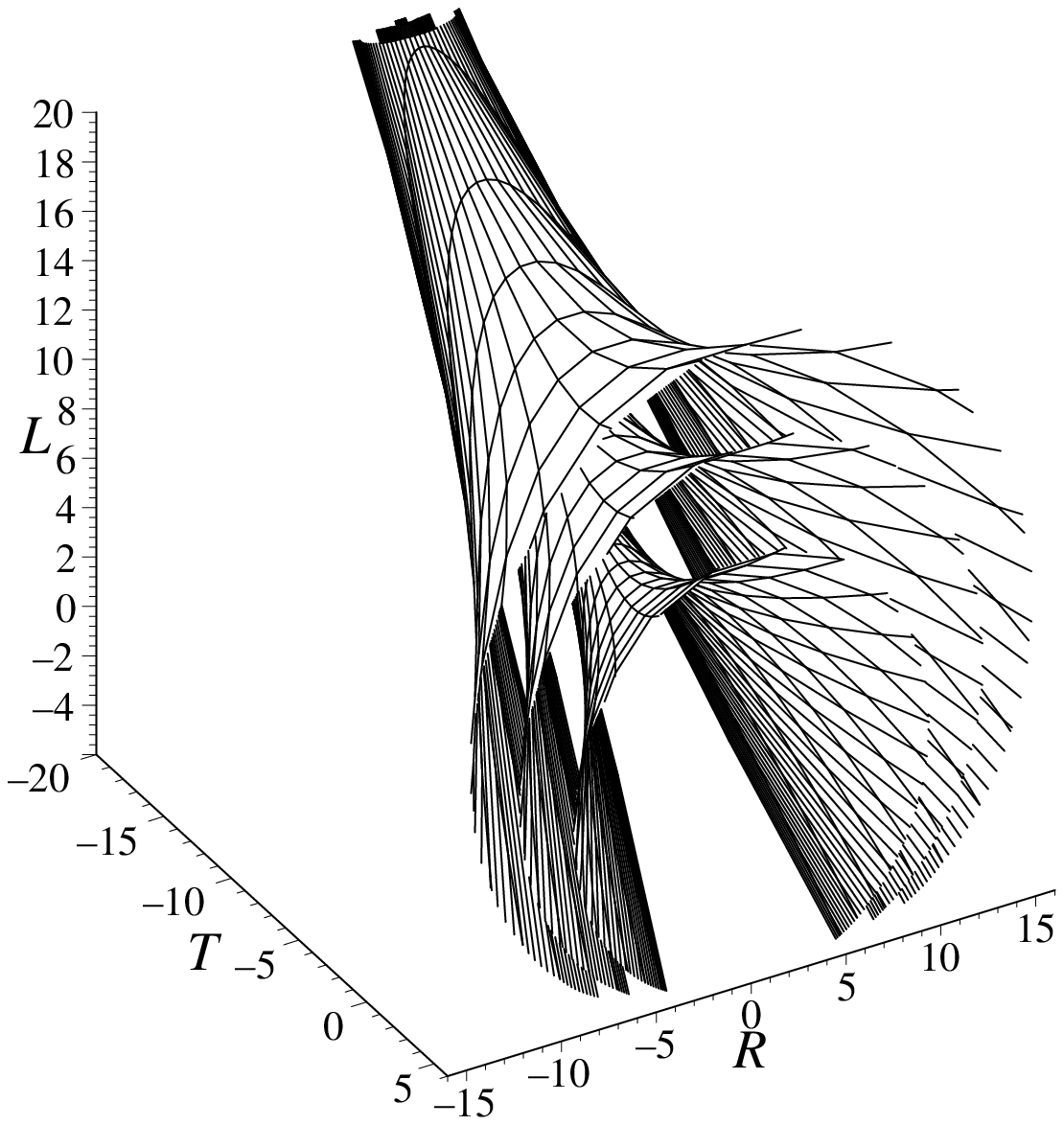}}
\caption{(left) Solution with $\alpha=3/2$ and $\ell_{0}=1$;
(right) Nest of solutions with $\alpha=1/3$ and $\ell_{0}=20,40,60$.}
\end{figure}

Embeddings can be approached in two ways, depending on whether theorems of differential
geometry are used to obtain generic results, or whether solutions of differential
equations are used which encode physics.  Here we adopt the latter approach, and
concentrate on two models which are believed to be relevant to the real universe.
Both involve a 5D flat space with simple coordinates $T,R,L$ which label time, radius,
and extension in the extra dimension.  In this manifold we embed 4D curved spaces
with matter, in which the corresponding coordinates $t,r$ are the ones used in
observational cosmology.  (The big manifold always has one extra coordinate compared to
the little manifold, the two being related by values of $\ell=\ell_{0}$ as noted above.)
The evolution of the models in 4D depends on the nature of the matter they contain,
and we illustrate two cases in Fig. 1.  The matter is described by a parameter
$\alpha$ which appears in the Ponce de Leon class of solutions.  In Fig. 1 
(left), $\alpha=3/2$
corresponds to an equation of state where there is no pressure between the galaxies as
in dust and we have set $\ell_{0}=1$.  In Fig. 1 (right), $\alpha=1/3$ corresponds to a force
that pushes particles rapidly apart as for inflation, and we have set $\ell_{0}=20,40,60$
to obtain a nest of solutions.  Thus Fig. 1 represent the present and early
universe.  The important thing, apart from their illustrative value, is that these plots
show the big bang in 4D to be a result of the embedding.  The absence of an initial
singularity in the flat embedding space is of great philosophical importance; and on a
technical level, results like those summarized here have significant implications for the
question of whether extra dimensions are merely convenient mathematical devices or are
real in a physical sense.

\section{Conclusion}
We have in the foregoing sections outlined the main astrophysical implications of
higher-dimensional gravity.  The latter subject exists in several forms, notably 5D
STM and membrane theories, 10D superstrings, 11D supergravity and 26D string
theory.  We have concentrated on the 5D case, because it is the basic extension of 4D
general relativity and the low-energy (nonperturbative) limit of higher-dimensional
theories.  It also has the practical advantage of allowing calculations to be done which
are open to test.  Among testable consequences, we have noted the change in the precession
of a gyroscope aboard a satellite in Earth orbit, and the perturbations due to solitons in
the cosmic microwave background.

In closing, we wish to remind our readers that these and other effects of higher dimensions
are subtle.  This is the nature of the subject.  When Newton's theory involving separate
time and space was extended by Einstein to general relativity, where the coordinates are
taken on the same footing and in fact mixable through covariance, the resulting observable
differences were minor.  Nevertheless, they were eventually confirmed by astrophysics.
The current situation is similar.  Higher-dimensional field theory seeks to unify gravity
with the interactions of particle physics, but adheres to the tradition that the new should
incorporate the old.  This necessarily means that the observational tests of new theory
involve subtle differences.  We have focussed on 5D because it appears to offer the best
prospect for tests at the current stage. (No value of $N$ is sacrosanct: the extension from
4D to 5D by Kaluza and Klein has been followed by the other theories mentioned above, and
indeed Kalitzin and others have considered $N\rightarrow\infty$, which however implies even
smaller modifications with a lower likelihood of verification.)  It seems to us that in the
near future effort should be directed at separating the two versions of 5D relativity.
In the STM approach, the effects of the extra dimension are all around us in the form of
energy, and the five dimensions are treated equally.  In membrane theory, gravity propagates
freely in the ``bulk'' of the extra dimension, whereas the other interactions are confined
to the hypersurface we call spacetime.    The mathematics of these two approaches was
recently shown by Ponce de Leon and others to be equivalent, but the physics is different
and needs to be investigated through astrophysical observation.

A logical area of investigation, beyond those mentioned above, is dark matter.  We know
(from data on galaxies, clusters and gravitational lensing) that this exists.  But its
nature is presently unknown.  We suggest that effort be directed at working out the
equation of state and other characteristics of dark matter on the basis of the various
versions of higher-dimensional gravity (and specifically STM and membrane theory).
A well-defined prediction from theory should with time be open to test from observation.

\section*{Acknowledgments}
We thank J. Ponce de Leon, S.S. Seahra, and H.R. Sepangi for useful comments and discussions.
We are grateful to R.C. Myers for suggestions on an earlier version of the manuscript.  This
work was supported by N.S.E.R.C.

\appendix

\section{Conventions}
In this paper 4D quantities are labelled with lower-case Greek indices that run from $0,\ldots,3$
while 5D ones are labelled using upper-case Latin indices that run from $0,\ldots,4$.  The
signature of the metric is always $(+-...-)$.  A convenient choice of coordinates is $x^{0}=t$,
$x^{123}=r\theta\phi$ $(d\Omega^{2}\equiv d\theta^{2}+\sin^{2}\theta d\phi^{2})$ with
$x^{4}=\ell$ except in $\S$VI where we use $x^{4}=y$.  The 5D covariant derivative is denoted
$\nabla_{A}$, and we use an asterisk $(\stackrel{*}{\phantom{p}})$ to denote
$\partial/\partial\ell$ where convenient.

\section*{References}
\begin{enumerate}
\item
Belayev, W.B. 2001, \emph{Spacetime and Substance}, \textbf{7}, 63.
\item
Billyard, A.P., Wesson, P.S. 1996a, \emph{Phys. Rev. D}. \textbf{53}, 731.
\item
Billyard, A.P., Wesson, P.S. 1996b, \emph{Gen. Rel. Grav}. \textbf{28}, 129.
\item
Billyard, A.P., Sajko, W.N. 2001, \emph{Gen. Rel. Grav}. \textbf{33}, 1929.
\item
Campbell, J.E. 1926, \emph{A Course of Differential Geometry}, Clarendon, Oxford.
\item
Chamblin, A. 2001, \emph{Class. Quant. Grav.} \textbf{18}, L17.
\item
Chiba, M., Yoshii, Y. 1999, \emph{Astrophys. J.} \textbf{510}, 42.
\item
Davidson, A., Owen, D. 1985, \emph{Phys. Lett. B}. \textbf{155}, 247.
\item
Eppley, J.M., Partridge, R.B. 2000, \emph{Astrophys. J.} \textbf{538}, 489.
\item
Eubanks, T. M. et al. 1997, \emph{Bull. Am. Phys. Soc.} \textbf{K11.05}
\item
Gibbons, G.W. 1982, \emph{Nucl. Phys. B.} \textbf{207}, 337.
\item
Green, M.B., Schwarz, J.H., Witten, E. 1987, \emph{Superstring Theory}, Camb. Univ. Press,
Cambridge.
\item
Gross, D. J., Perry, M.J. 1983, \emph{Nucl. Phys. B}. \textbf{226}, 29.
\item
Halpern, P. 2001, \emph{Phys. Rev. D}. \textbf{63}, 024009.
\item
Hoyle, F., Narlikar, J.V. 1974, \emph{Action at a Distance in Physics and Cosmology},
Freeman, San Francisco.
\item
Kalligas, D., Wesson, P.S., Everitt, C.W.F. 1995, \emph{Astrophys. J}. \textbf{439}, 548.
\item
Lidsey, J.E., Romero, C., Tavakol, R., Rippl, S. 1997,
\emph{Class. Quant. Grav.} \textbf{14}, 865.
\item
Lineweaver, C.H. 1998, \emph{Astrophys. J.} \textbf{505}, L69.
\item
Liu, H., Mashhoon, B. 2000, \emph{Phys. Lett. A.} \textbf{272}, 26.
\item
Liu, H., Overduin, J.M. 2000, \emph{Astrophys. J.} \textbf{538}, 386.
\item
Maartens, R. 2000, \emph{Phys. Rev. D.} \textbf{62}, 084023.
\item
Mansouri, F. 2002, \emph{Phys. Lett. B}. \textbf{538}, 239.
\item
Mashhoon, B., Liu, H., Wesson, P.S. 1994, \emph{Phys. Lett. B}. \textbf{331}, 305.
\item
Mashhoon, B., Liu, H. 2000, \emph{Phys. Lett. A}. \textbf{272}, 26.
\item
Matos, T. 1987, \emph{Gen. Rel. Grav.} \textbf{19}, 481.
\item
Overduin, J.M., Wesson, P.S. 1997, \emph{Phys. Rep}. \textbf{283}, 303.
\item
Overduin, J.M., Cooperstock, F.I. 1998, \emph{Phys. Rev. D.} \textbf{58}, 043506.
\item
Overduin, J.M. 1999, \emph{Astrophys. J.} \textbf{517}, L1.
\item
Overduin, J.M. 2000, \emph{Phys. Rev. D.} \textbf{62}, 102001.
\item
Overduin, J.M., Wesson, P.S. 2003, \emph{Dark Sky, Dark Matter}, IOP, Bristol.
\item
Ponce de Leon, J. 1988, \emph{Gen. Rel. Grav.} \textbf{20}, 539.
\item
Ponce de Leon, J. 2001, \emph{Phys. Lett. B}. \textbf{523}, 311.
\item
Ponce de Leon, J. 2002, \emph{Int. J. Mod. Phys.} \textbf{11}, 1355.
\item
Ponce de Leon, J. 2003a, \emph{Int. J. Mod. Phys.} \textbf{12}, 757.
\item
Ponce de Leon, J. 2003b, \emph{Gen. Rel. Grav.} \textbf{35}, 1363.
\item
Rippl, S., Romero, C., Tavakol, R. 1995, \emph{Class. Quant. Grav.} \textbf{12}, 2411.
\item
Romero, C., Tavakol, R., Zalatetdinov, R. 1996, \emph{Gen. Rel. Grav}. \textbf{28}, 365.
\item
Seahra, S.S., Wesson, P.S. 2001, \emph{Gen. Rel. Grav}. \textbf{33}, 1752.
\item
Seahra, S.S. 2002, \emph{Phys. Rev. D.} \textbf{65}, 124004.
\item
Seahra, S.S., Wesson, P.S. 2002, \emph{Class. Quant. Grav.} \textbf{19}, 1139.
\item
Seiler, W.M., Roque, W.L. 1991, \emph{Gen. Rel. Grav}. \textbf{23}, 1151.
\item
Sorkin, R.D. 1983, \emph{Phys. Rev. Lett.} \textbf{51}, 87.
\item
Weinberg, S. 1972, \emph{Gravitation and Cosmology}, Wiley, New York.
\item
Weinberg, S. 2000, \emph{The Quantum Theory of Fields III: Supersymmetry}, Camb. Univ. Press,
Cambridge.
\item
Wesson, P.S. 1978, \emph{Cosmology and Geophysics}, Hilger/Oxford Univ. Press, New York.
\item
Wesson, P.S. 1992, \textit{Astrophys. J}. \textbf{394}, 19.
\item
Wesson, P.S. 1994, \emph{Astrophys. J.} \textbf{420}, L49.
\item
Wesson, P.S., Liu, H. 1998, \emph{Phys. Lett. B.} \textbf{432}, 266.
\item
Wesson, P.S. 1999, \emph{Space-Time-Matter}, World Scientific, Singapore.
\item
Wesson, P.S., Mashhoon, B., Liu, H., Sajko, W.N. 1999, \emph{Phys. Lett B}.
\textbf{456}, 34.
\item
Wesson, P.S., Liu, H. 2001, \emph{Int. J. Mod. Phys. D.} \textbf{10}, 905.
\item
Wesson, P.S. 2002a, \emph{Class. Quant. Grav}. \textbf{19}, 2825.
\item
Wesson, P.S. 2002b, \emph{J. Math. Phys}. \textbf{43}, 2423.
\item
Will, C.M. 1993, \emph{Theory and Experiment in Gravitational Physics}, Camb. Univ. Press,
Cambridge.
\item
Youm, D. 2000, \emph{Phys. Rev. D.} \textbf{62}, 084002.
\item
Youm, D. 2001, \emph{Mod. Phys. Lett. A.} \textbf{16}, 2371.
\end{enumerate}

\end{document}